\documentstyle{article}

    \font\cmss=cmss10
    
         \def\ba{\begin{array}}
         \def\ea{\end{array}}
         \def\be{\begin{equation}}
         \def\ee{\end{equation}}
         \def\det{{\rm det}}
         
         \def\SL{{\rm SL}}

         \def\sl{\hbox{{\cmss sl}}}
         \def\half{{1 \over 2}}
         
         \def\SU{{\rm SU}}
         
         \def\Uh#1{{\rm U}_h \left( #1 \right) }
         \def\Um#1{{\rm U}_{\mu} \left( #1 \right) }
         \def\U#1{{\rm U} \left( #1 \right) }
         
         \def\R{{\rm R}}
         \def\M{{\rm M}}
         \def\dM{{\rm dM}}
         \def\Real{\hbox{{\cmss R}}}

         \def\Poi#1{\hbox{{\cmss P}}(#1)}
         \def\poi#1{\hbox{{\cmss p}}(#1)}
         
         \def\A{{\alpha}}
         \def\B{{\beta}}
         \def\C{{\gamma}}
         \def\D{{\delta}}
         \def\d{{\rm d}}
         \def\w{{\wedge}}
         \def\um{{ 1 \over \mu }}
\begin{document}
%%%%%%%%%%%%%%%%%%%%  The title page   %%%%%%%%%%%%%%%%%%%%%%%%%%%%%%%%%%%%%
\hfill
\vbox{
    \halign{#\hfil         \cr
            hep-th/9412029 \cr
            IPM-94-071     \cr
            IASBS 94-6     \cr
            TUDP 94-2      \cr
            SUTDP 94-73-11 \cr
           } % end of \halign
      }  % end of \vbox
\vskip 10 mm
\leftline{ \Large \bf
          A Triangular Deformation of the }
\leftline{ \Large \bf
           two Dimensional Poincar\'e Algebra }
\vskip 10 mm
\rm
\leftline{ \bf
          M. Khorrami ${}^{1,2,3,*}$,
          A. Shariati ${}^{1,4}$,
          M. Abolhasani ${}^{1,4}$,
          A. Aghamohammadi ${}^{1,5}$
          }
\vskip 10 mm
{\it
  \leftline{ $^1$ Institute for Studies in Theoretical Physics and
            Mathematics, P.O.Box  5746, Tehran 19395, Iran. }
  \leftline{ $^2$ Department of Physics, Tehran University,
             North-Kargar Ave. Tehran, Iran. }
  \leftline{ $^3$ Institute for Advanced Studies in Basic Sciences,
             P.O.Box 159, Gava Zang, Zanjan 45195, Iran. }
  \leftline{ $^4$ Department of Physics, Sharif University of Technology,
             P.O.Box 9161, Tehran 11365, Iran. }
  \leftline{ $^5$ Department of Physics, Alzahra University,
             Tehran 19834, Iran }
  \leftline{ $^*$ E-Mail: mamwad@irearn.bitnet}
  }
\begin{abstract}
Contracting the  $h$-deformation of $\SL(2,\Real)$, we construct a
new deformation of two dimensional Poincar\'e algebra, the algebra of
functions on its group and its differential structure. It is also shown that
the Hopf algebra is triangular, and its universal R matrix is also
constructed explicitly. Then, we find a deformation map for the universal
enveloping algebra, and at the end, give the deformed mass shells and Lorentz
transformation.
\end{abstract}
\vskip 10 mm
%%%%%%%%%%%%%%%%%%%%  The body of the paper            %%%%%%%%%%%%%%%%%%%%%
\section{ Introduction }
Deformations of the Poincar\'e  group have recieved considerable interest
in recent years [1-3].
These deformations may be considered as the
authomorphisms of the quantum deformed space-time. Although a general theory
of quantum deformations of inhomogenous quantum groups is lacking, there are
attempts for constructing these non-semisimple groups [1-5].
As we know, there are at least three distinct deformations of the Poincar\'e
group in four dimensions. The first is $q$-$\Poi{4}$, introduced by
Ogievetsky et al. \cite{Og}.
The second one is $\kappa$-$\Poi{4}$, introduced by Lukierski et al.
\cite{LNRT,LN}.
This $\kappa$-$\Poi{4}$ is obtained by a contraction method. There exist
another deformation which is constructed in \cite{Chai}.
A natural question to ask, is the number of distinct deformations of
$\Poi{4}$.

It seems easier to consider the two dimensional case. In fact, the first step
for constructing inhomogeneous  quantum groups is related to the $q$-analog
of two dimensional Euclidean group [7-9].
There exist two distinct deformations of E(2), both of which can be
obtained from $\SU_q(2)$ by a contraction procedure \cite{Vaks,CGST,BCGST};
neither of them has a universal R mtrix.
In two dimensions, however, there exist two distinct deformations of
$\SL(2)$, the well-known $q$-deformation, and the $h$-deformation [11-19].
The $h$-deformation itself can be obtained from the $q$-deformation
by a contraction procedure \cite{AKS}.

In this article we obtain a new deformation of $\Poi{2}$ from the contraction
of the $h$-deformation of $\SL(2,\Real)$. First, we construct
the deformed universal enveloping algebra $\Um{\poi{2}}$, the
deformed algebra
of functions on the $\Poi{2}$, and its differential calculus.
Then, we show that $\Um{\poi{2}}$ has a universal R matrix, and
explicitly construct the universal R matrix. It is seen that this Hopf
algebra is triangular. This is the distinguishing feature of this
deformation of $\Poi{2}$, note that
the other deformations of E(2) don't have universal R matrices.
We also give a deformation map which relates
the commutation relations of $\Um{\poi{2}}$ to their classical counterparts.
This can be used to study the representation theory of $\Um{\poi{2}}$.
Finally, we give the deformed mass shells and finte Lorentz transformation
for energy and momentum in light cone coordinates.

\section{ $\mu$-Deformation of the two dimensional Poincar\'e group }

The $h$-deformation of $\SL(2,\Real)$ has the following structure. The deformed
algebra of functions, $\SL_h(2)$ is generated by the enteries of the matrix
\be
   \M := \pmatrix{ a & b \cr c & d \cr }.
\ee
and has the following Hopf algebra structure.
\be
 \ba{lll}
  [a,c]=hc^2                        &
  [d,b]=h(D-d^2)                    &
  [a,d]=h(d c- ac)                  \cr
  [d,c]=h c^2                       &
  [b,c] =h( a c +  c d)             &
  [b,a] = h (a^2 -D)                \cr
 \ea
\ee
\be
 \ba{ll}
  \Delta a=a\otimes a+b\otimes c    &
  \Delta b=a\otimes b+b\otimes d    \cr
  \Delta c=c\otimes a+d\otimes c    &
  \Delta d=c\otimes b+d\otimes d    \cr
 \ea
\ee
\be
 \ba{ll}
  S(a)=d + h c                      &
  S(b)=-b-h (a-d) +h^2 c            \cr
  S(c)=-c                           &
  S(d)=a-hc                         \cr
\ea
\ee
\be
 \ba{ll}
  \epsilon (a)=1                    &
  \epsilon (b)=0                    \cr
  \epsilon (c)=0                    &
  \epsilon (d)=1                    \cr
 \ea
\ee
\be
 D = \det_h \M := a \, d - c \, b - h \, c \, d=1
\ee
The deformed universal enveloping algebra is generated by three generators
$X$, $Y$, and $H$. The Hopf algebra structure, as first introduced by Ohn
\cite{Ohn}, is the following.
\be
 \ba{l}
  [ H , X ] = \displaystyle{ 2 \over h} \sinh hX   \cr
  [ H , Y ] = - \{ Y , \cosh h X \}  \cr
  [ X , Y ] = H                       \cr
 \ea
\ee
\be
 \ba{l}
  \Delta X=X\otimes 1+1\otimes X  \cr
  \Delta Y=Y\otimes {\rm exp}(hX) + {\rm exp}(-hX)\otimes Y \cr
  \Delta H=H\otimes {\rm exp}(hX)+{\rm exp}(-hX)\otimes H   \cr
 \ea
\ee
\be
 \ba{l}
  S(X)=-X                                \cr
  S(Y)=-{\rm exp}(hX) \,Y \, {\rm exp}(-hX)  \cr
  S(H)=-{\rm exp}(hX)\, H \, {\rm exp}(-hX)  \cr
 \ea
\ee
\be
 \ba{lll}
  \epsilon (X) = 0    &
  \epsilon (Y) = 0    &
  \epsilon (H) = 0    \cr
 \ea
\ee
These two Hopf algebras are dual. The duality is
\be \label{duality}
  \ba{l}
  \Big<H,  \pmatrix{ a & b \cr c & d \cr }\Big>=
   \pmatrix{ 1 & 0 \cr 0 & -1 \cr } \cr\cr
  \Big<Y,  \pmatrix{ a & b \cr c & d \cr }\Big>=
   \pmatrix{ 0 & 0 \cr 1 & 0 \cr } \cr\cr
  \Big<X,  \pmatrix{ a & b \cr c & d \cr }\Big>=
   \pmatrix{ 0 & 1 \cr 0 & 0 \cr } \cr
  \ea
\ee
Now we use the following contraction to obtain a new deformation of $\Poi{2}$
and $\U{\poi{2}}$. For $\Uh{\sl(2,\Real)}$ we define
\be
  P^+:=\epsilon X,\qquad P^-:=\epsilon Y \qquad K:=\half H \qquad
\mu:=\epsilon h^{-1}
\ee
As we shall see, $P^{\pm}$, are the deformed momenta in the light cone
coordinates, and $K$ is the deformed Lorentz boost.
In the $\epsilon \to 0$ limit we get the following Hopf algebra which is
a deformation of $U(\Poi{2})$.
\be \label{CR}
 \ba{l}
  [ K , P^+ ] = \mu \sinh ({P^+\over \mu})   \cr
  [ K , P^- ] = - P^-  \cosh({P^+\over \mu}) \cr
  [ P^+, P^- ] =0                             \cr
 \ea
\ee
\be
 \ba{l}
   \Delta P^+ = P^+\otimes 1+1\otimes P^+                \cr
   \Delta P^- = P^-\otimes {\rm exp}({P^+\over \mu}) +
                 {\rm exp}(-{P^+\over \mu})\otimes P^-    \cr
   \Delta K   = K\otimes {\rm exp}({P^+\over \mu})+
                 {\rm exp}({P^+\over \mu})\otimes K       \cr
 \ea
\ee
\be
 \ba{lll}
  S(P^+) = - P^+          &
  S(P^-) = - P^-          &
  S(K)   = - K            \cr
  \epsilon (P^+) = 0       &
  \epsilon (P^-) = 0       &
  \epsilon (K)   = 0      \cr
 \ea
\ee
To obtain the deformed algebra of functions, we use the known duality
(\ref{duality}), from which we conclude that the corresponding
contraction in the algebra of functions is
\be
  \A:=a \qquad \B:=\epsilon^{-1} b,\qquad \C:=\epsilon^{-1} c \qquad
  \D := d \qquad \mu:=\epsilon h^{-1}.
\ee
Expressing the commutation relations of $\SL_h(2,\Real)$ in terms of
these new generators, and going to the $\epsilon \to 0$ limit, we obtain
\be
  \ba{ll}
   [\A,\B]=\mu^{-1}(1-\A^2)  &
   [\A,\D]= 0                \cr
   [\D,\B]=\mu^{-1}(1-\D^2)  &
   [\C,\cdots ] = 0          \cr
  \ea
\ee
The four generators of $\SL_h(2,\Real)$ are not independent since the
determinent of M must be equal to one. This leads to
\be
   {\rm det}_{\mu} \M= \A \D=1 \qquad \Rightarrow \qquad \D = \A^{-1}
\ee
The calculations for the co-product, co-unity, and antipode is
straightforward. For completeness, we write them here.
\be
  \ba{ll}
   \Delta \A=\A\otimes \A                    &
   \Delta \B=\A\otimes \B+\B\otimes \D      \cr
   \Delta \C=\C\otimes \A+\D\otimes \C       &
   \Delta \D=\D\otimes \D                   \cr
  \ea
\ee
\be
 \ba{ll}
  S(\A)=\D                                  &
  S(\B)=-\B-\mu^{-1} (\A-\D)               \cr
  S(\C)=-\C                                 &
  S(\D)=\A                                 \cr
 \ea
\ee
\be
 \ba{ll}
  \epsilon (\A)=1                           &
  \epsilon (\B)=0                           \cr
  \epsilon (\C)=0                            &
  \epsilon (\D)=1                           \cr
 \ea
\ee

The differential structure of {\cmss P}$_{\mu}(2)$ can be obtained in the
similar way. One can use differential structure of $\SL_h(2)$, applying the
same contraction proccess in the level of differentials too. To construct
differential structure one should use the following relations
\cite{Sud,SWZ,Kar}
\be \R\dM_1\M_2=\M_2\dM_1\R  \ee
\be \label{2} \R\dM_1 \wedge \dM_2=-\dM_2\wedge \dM_1\R, \ee
where
\be
  \ba{ll}
  \R = \pmatrix{
                1 & -h & h & h^2 \cr
                0 &  1 & 0 & -h  \cr
                0 &  0 & 1 &  h  \cr
                0 &  0 & 0 &  1  \cr
               }
  &
  dM = \pmatrix{ \d a & \d b \cr \d c & \d d \cr }.
  \cr
 \ea
\ee
After applying the contraction, for the algebra of $M_{ij}$ and $\d M_{ij}$
we obtain:
\be
 \ba{ll}
  \B \d \C = \d \C \B + \um ( \D + \A ) \d \C     &
  \C \d \B = \d \B \C - \um \C ( \d \D + \d \A )  \cr
  \B \d \A = \d \A \B + \um ( \A - \D ) \d \A     &
  \A \d \B = \d \B \A + \um \A ( -\d \A + \d \D ) \cr
  \B \d \D = \d \D \B + \um ( \D - \A) \d \D      &
  \D \d \B = \d \B \D + \um \D ( \d \A - \d \D )  \cr
  \B \d \B = \d \B \B + (\um)^2 ( \D \d \D - \A \d \A ) &
  \hskip -9pt + \um ( \D \d \B - \B \d \D + \B \d \A - \A \d \B ) \cr
 \ea
\ee
All the remaining relations are commutative.
After the contraction (\ref{2}), results in;
\be
\ba{ll}
\d \A \w \d \A = \d \C \w \d \C = \d \D \w \d \D = 0 &
\d \B \w \d \B = { 1 \over 2 \mu} \
\left\{ \d \B \w \left( \d \D + \d \A \right) - \left( \d \D + \d \A \right) \w
\d \B \right\} \cr
\d \A \w \d \B + \d \B \w \d \A = \um \d \A \w \d \D &
\d \B \w \d \C + \d \C \w \d \B = \um (\d \D \w \d \C - \d \C \w \d \A ) \cr
\d \B \w \d \D + \d \D \w \d \B = \um \d \D \w \d \A &
\d \A \w \d \C + \d \C \w \d \A = 0 \cr
\d \D \w \d \C + \d \C \w \d \D = 0 &
\d \A \w \d \D + \d \D \w \d \A = 0 \cr
\ea \ee
%---------------------------------------------------------------------------
\section{ Universal R matrix}
A quasitriangular Hopf algebra is a Hopf algebra with a universal R matrix
satisfying
\be \label{quasitri1}
\ba{ll}
( \Delta \otimes 1 ) R = R_{13} R_{23} \cr\cr
( 1 \otimes \Delta ) R = R_{13} R_{12}  \cr
\ea \ee
\be \label{quasitri2}
R \Delta (\cdot ) R^{-1} = \Delta' ( \cdot ) := \sigma \circ \Delta ( \cdot )
\ee
where $\sigma$ is the flip map: $\sigma (a \otimes b) = b \otimes a $.
If in addition
\be \label{triangular}
\sigma ( R^{-1} ) = R
\ee
the Hopf algebra is called  triangular \cite{Majid}.

Vladimirov \cite{Vlad} has considered the subalgebra of $\Uh{\sl(2)}$ generated
by the two generators $X$ and $H$ and has found the following universal R
matrix for this subalgebra.
\be \label{RVlad}
\tilde{R} = \exp \left\{ { {h \Delta X} \over \sinh (h \Delta X ) }
\left[ H \otimes \sinh ( h X ) - \sinh (h X ) \otimes H \right] \right\}
\ee
Expressing this R matrix in terms of the new generators, and going to the
$ \epsilon \to 0 $ limit, we get the following R matrix.
\be \label{Rmatrix}
\ba{ll}
\displaystyle{ R = \exp (A) } &
\displaystyle{ A= 2 { { \Delta \pi^+} \over  \sinh ( \Delta \pi^+ ) }
\left( K \otimes \sinh \pi^+ - \sinh  \pi^+ \otimes K \right).} \cr\ea \ee
We have used the following notation.
\be
\pi^{\pm}:=\mu^{-1}P^{\pm}
\ee
We will show that this is the universal R matrix for $\Um{\Poi{2}}$. First, it
is seen that this matrix is not singular. Second, as  Vladimirov's R matrix,
(\ref{RVlad}), satisfies the quantum Yang-Baxter equation, the relations
(\ref{quasitri1}), and also
\be \ba{l}
\tilde{R} \Delta  X  \tilde{R}^{-1} = \Delta' X \cr
\tilde{R} \Delta  H  \tilde{R}^{-1} = \Delta' H, \cr
\ea \ee
it is obvious that this new R matrix also satisfies the quantum Yang-Baxter
equation, the relations (\ref{quasitri1}), and also
\be \ba{l}
R \Delta  \pi^+ R^{-1} = \Delta' \pi^+ \cr
R \Delta  K   R^{-1} = \Delta' K. \cr
\ea \ee
Now we want to prove that $R$ is indeed the universal R matrix of the whole
algebra $\Um{\Poi{2}}$.

Define
\be
E := R \Delta P^- R^{-1} - \Delta' P^- . \ee
We are to prove that $E = 0$. To see this we use the commutation relation of
$P^-$ and $K$, and the fact that $\Delta$ and $\Delta'$ are homomorphisms of
the algebra:
\be \ba{ll}
\left[ R \Delta P^- R^{-1} , \Delta' K \right]\hskip -9pt &=  R \Delta
P^- R^{-1} \cosh (\Delta'\pi^+) \cr
&= (E+\Delta'P^-)\cosh (\Delta'\pi^+)\cr
\ea \ee
and,
\be \ba{ll}
\left[ R \Delta P^- R^{-1} , \Delta' K \right]\hskip -9pt &=
\left[ E + \Delta' P^- , \Delta' K \right] \cr
&= \left[ E, \Delta'K \right] + \Delta' P^- \cosh (\Delta'\pi^+) \cr
\ea \ee
which yields
\be
\left[ \Delta' K , E \right] = - E \cosh \Delta' \pi^+ . \ee
On the other hand, using Hausdorf's identity, we have
\be
R \Delta P^- R^{-1} = \exp \left( \left[ A , \cdot \, \right] \right) \Delta
P^-
\ee
It is easy to see that the commutator of $A$ with a term like
$ P^- f(\pi^+) \otimes g(\pi^+) $ is a linear combination of other terms of
this kind. Moreover, if $f$ and $g$ are analytic functions around
$\pi^+ = 0 $, the commutator is also analytic around $\pi^+ = 0$. Using
these, one can write
\be
E = C + D \ee
where
\be
\ba{l}
 C:= \displaystyle{\sum_{m,n \ge 0} C_{mn} P^- ( \pi^+ )^m \otimes
 ( \pi^+ )^n} \cr\cr
 D:= \displaystyle{\sum_{m,n \ge 0} D_{mn}  ( \pi^+ )^m \otimes P^-
 ( \pi^+ )^n}, \cr\cr
\ea \ee
and
\be \ba{l}
\left[ \Delta' K , C \right] = - C \cosh \left( \Delta' \pi^+ \right) =
- \left( C \cosh \pi^+ \otimes \cosh \pi^+ + C \sinh \pi^+ \otimes \sinh \pi^+
\right) \cr
\left[ \Delta' K , D \right] = - D \cosh \left( \Delta' \pi^+ \right) =
- \left( D \cosh \pi^+ \otimes \cosh \pi^+ + D \sinh \pi^+ \otimes \sinh \pi^+
\right). \cr
\ea \ee
The first relation leads to
\be \label{gonde} \ba{l}
\displaystyle{\sum_{m,n \ge 0} C_{mn} \left[ \left( m { { \sinh \pi^+ }
\over \pi^+ } -\cosh \pi^+ \right) P^- ( \pi^+ )^m \otimes ( \pi^+ )^n
\exp ( -\pi^+ ) +n P^- ( \pi^+ )^m \exp ( \pi^+) \otimes ( \pi^+ )^n
{{ \sinh \pi^+ }  \over\pi^+} \right]} \cr \cr
 = - \displaystyle{\sum_{m,n \ge 0} C_{mn} \left[
   P^- ( \pi^+ )^m \cosh \pi^+ \otimes ( \pi^+ )^n \cosh \pi^+ +
 P^- ( \pi^+ )^m \sinh \pi^+ \otimes ( \pi^+ )^n  \sinh \pi^+
\right]} \cr
\ea \ee
Now, suppose that at least one of the $C_{mn}$'s is nonzero. Then, it is easy
to see that there exists a pair $(m_0 , n_0 )$ such that
\be \ba{l} C_{m_0 \, n_0} \ne 0 \cr
C_{mn} = 0 \hskip 10 mm {\rm if} \hskip 10 mm
\big( m \le m_0 , n \le n_0 , ( m,n) \ne (m_0 , n_0 ) \big) \cr
\ea \ee
Expanding the relation (\ref{gonde}) in powers of $\pi^+$, and equating the
coefficients of $P^- ( \pi^+ )^{m_0} \otimes ( \pi^+ )^{n_0} $ on both sides
of the equation, one arrives at
\be ( m_0 - 1 + n_0 ) C_{m_0 \, n_0 }  = - C_{m_0 \, n_0 } \ee
or,
\be ( m_0 + n_0 ) C_{m_0 \, n_0 } = 0, \ee
which  yields
\be m_0 = n_0 = 0. \ee
So, if one can show that $C_{00} = 0$, it implies that $ C = 0 $. The same
argument is also true for $D$. this means that if $E$ is zero up to zeroth
order of $\pi^+$, it is zero to all orders of $\pi^+$. But we have
\be R \vert_{\pi^+ = 0} =1 \hskip 10 mm \Rightarrow \hskip 10 mm
R \Delta P^- R^{-1} \vert_{\pi^+ = 0} = 1 \otimes P^- + P^- \otimes 1 =
\Delta' P^- \vert_{\pi^+ = 0} \hskip 10 mm \Rightarrow \ee
\be E \vert_{\pi^+ = 0} = 0 \ee
This completes the proof that this Hopf algebra is quasitriangular,
\be R \Delta P^- R^{-1} = \Delta' P^- .\ee
Now it can be easily shown that (\ref{triangular}) is indeed satisfied for
the R matrix (\ref{Rmatrix}). This shows that $\Um{\Poi{2}}$ is triangular.

\section{ Deformation map}

Following the idea of \cite{CZ}
one can find a deformation map for the $\Um{\poi{2}}$ which reads as
follows:
\be
\ba{l}
\pi^+ := \widehat{\pi}^+ \cr\cr
\pi^- := \displaystyle{{ \sinh  \widehat{\pi}^+ \over \widehat{\pi}^+}}
         \widehat{\pi}^- \cr\cr
K:= \displaystyle{{\widehat{\pi}^+ \over \sinh \widehat{\pi}^+ }}
         \widehat{K} \cr
\ea \ee
where $\widehat\pi^{\pm}$ and $\widehat K$ belong to the classical, i.e.
undeformed, algebra. A simple calculation shows that the commutation
relations (\ref{CR}) result from the commutation relations of the classical
case. However, the co-product structure are not related in this way. This
deformation map can be used to study the representation theory of
$\Um{\poi{2}}$ as in the case of $\kappa$-Poincar\'e algebra \cite{Mas}.
A furthure remark is that, as in the
case of $\kappa$-Poincar\'e algebra, the deformation map is
singular for certain values of $P^+$ if $\mu$ is imaginary, because in this
case the hyperbolic functions are replaced by trigonometric ones.

\section{ Mass-shells and finite boosts }

The casimir of $\Um{\poi{2}}$ is
\be
  C_2 := \pi^- \sinh \pi^+
\ee
To obtain finite boosts we postulate that $K$ transforms any operator
according to
\be
 {\d \Omega \over \d \eta } = [ K , \Omega ].
\ee
One can solve the system of differential equations
\be
 \ba{l}
  \displaystyle{ \d \pi^+ \over \d \eta }
  = [ K , \pi^+ ] = \sinh \pi^+       \cr\cr
  \displaystyle{ \d \pi^- \over \d \eta }
  = [ K , \pi^- ] = - \pi^- \cosh \pi^+ \cr
 \ea
\ee
and get
\be
 \ba{l}
 \tanh \pi^+ = \exp(\eta) \tanh \pi^+_0 \cr\cr
 ( \pi^- )^2 + \displaystyle{ m^2 \over \mu^2 } \pi^- =
 \exp( - 2 \eta ) \left[ (\pi^-_0)^2 +
 \displaystyle{ m^2 \over \mu^2 } \pi^-_0 \right],
 \ea
\ee
where $m$ is the {\it mass}, i.e. $C_2 = m^2$. As one expects, the
$ \mu \rightarrow \infty $ limit of these are the familiar Lorentz
transformation of energy and momentum in light cone coordinates.
For  imaginary $\mu$,  $\pi^+$ is an angle variable.
Figuers (1) and (2) show the mass-shells for real and imaginary $\mu$.
Note that, the topology of the light cone has changed in the case of
imaginary $\mu$.
One can compare this with the $\kappa$-deformation of the Poincar\'e
group \cite{LNRT}, the Casimir of which reads
\be
\displaystyle{
C_2 := 2 \kappa \sinh^2({P_0 \over 2 \kappa }) - P^2.
              }
\ee
Finite boosts for the $\kappa$-Poincar\'e are worked out in \cite{LRR}.
The result is something in terms of Jacobi elliptic functions. A point is
that, in four dimensional $\kappa$-Poincar\'e, one can set all the generators,
except $P_1$, $P_0$, and $L_1$, equal to zero, and arrive to the two
dimensional $\kappa$-Poincar\'e, which is the Lorentz counterpart of $E_l(2)$,
introduced in \cite{BCGST}. It may be possible to find a deformation of the
four
dimensional Poincar\'e group containing, in the same fashion, $\Um{\poi{2}}$.

%\begin{figure}
%  \vspace{100 mm}
%  \caption{ Mass-shells for $\mu$ real
%            }
%\end{figure}
%\begin{figure}
%  \vspace{100 mm}
%  \caption{ Mass-shells for $\mu = i \nu $ imaginary
%            }
%\end{figure}

%%%%%%%%%%%%%%%%%%%  References                       %%%%%%%%%%%%%%%%%%%%%

%%%%%%%%%%%%%%%%%%%%%%%%%%%%%%%%%%%%%%%%%%%%%%%%%%%%%%%%%%%%%%%%%%%%%%%%%%%%%%

\newpage
\begin{thebibliography}{99}
\bibitem{Og}
  O. Ogievetski, W. B. Schmidke, J. Wess, and B. Zumino; Lett. Mat. Phys.
  {\bf 23}, 233 (1991)
\bibitem{LNRT}
  J. Lukierski, A. Nowicki, H. Ruegg, and V. L. Tolstoy;
  Phys. Lett. {\bf B 264}, 331 (1991)
\bibitem{LN}
  J. Lukierski, and A. Nowicki; Phys. Lett. {\bf B 279}, 299 (1991)
\bibitem{Cas}
  L. Castellani; Phys. Lett. {\bf B279}, 291 (1992)
\bibitem{ARS}
  A. Aghamohammadi, S. Rouhani, and A. Shariati; to appear in J. Phys. A
\bibitem{Chai}
  M. Chaichian, and A. P. Demichev; Phys. Lett. {\bf B304}, 220 (1993)
\bibitem{Vaks}
  L. L. Vaksman, and L. I. Korogodski; Sov. Math. Dokl. {\bf 39}, 173 (1989)
\bibitem{Wor}
  S. L. Woronowicz; Lett. Math. Phys. {\bf 23}, 251 (1991)
\bibitem{CGST}
  E. Celeghini, R. Giachetti, E. Sorace, and M. Tarlini;
  J. Math. Phys. {\bf 32}, 1159 (1991)
\bibitem{BCGST}
  A. Ballesteros, E. Celeghini, R. Giachetti, E. Sorace, and M. Tarlini;
  J. Phys. {\bf A26}, 7495 (1993)
\bibitem{Dem}
  E. E. Demidov, Yu. I. Manin, E. E. Mukhin, and D. V. Zhdanovich;
  Prog. Theor. Phys. Suppl. {\bf 102}, 203 (1990)
\bibitem{Dub}
  M. Dubois-Violette, and G. Launer; Phys. Lett. {\bf B245}, 175 (1990)
\bibitem{Zak}
  S. Zakrzewski; Lett. Math. Phys. {\bf 22}, 287 (1991)
\bibitem{Ohn}
  Ch. Ohn; Lett. Math Phys. {\bf 25}, 85 (1992)
\bibitem{Kup}
  B. A. Kupershmidt; J. Phys. {\bf A25}, L1239 (1992)
\bibitem{Kar}
  V. Karimipour; Lett. Math. Phys. {\bf 30}, 87 (1994)
\bibitem{Agha}
  A. Aghamohammadi; Mod. Phys. Lett. {\bf A8}, 2607 (1993)
\bibitem{Vlad}
  A. A. Vladimirov; Mod. Phys. Lett. {\bf A8}, 2573 (1993)
\bibitem{AKS}
  A. Aghamohammadi, M. Khorrami, and A. Shariati; IPM-94-61, hep-th/9410135
\bibitem{Sud}
  A. Sudbery; Phys. Lett. {\bf B 284}, 61 (1992)
\bibitem{SWZ}
  P. Schupp, P. Watts, and B. Zumino; Lett. Math. Phys. {\bf 25}, 139 (1992)
\bibitem{Majid}
  S. Majid; Int. J. Mod. Phys. {\bf A5}, 1 (1990)
\bibitem{CZ}
  R. Curtright, C. Zachos; Phys. Lett. {\bf B243}, 237 (1990)
\bibitem{Mas}
  P. Ma\'slanka; J. Math. Phys. {\bf 34}, 6025 (1993)
\bibitem{LRR}
  J. Lukierski, H. Ruegg, and W. R\"uhl; Phys. Lett. {\bf B313}, 357 (1993)
\end{thebibliography}
\end{document}